\documentclass[a4paper,12pt]{amsart}
\usepackage{amsmath,amsfonts,amssymb,amscd,enumerate}

\def\A{\mathcal{A}}
\def\C{\mathbb{C}}

\def\R{\mathbb{R}}

\def\g{\mathfrak{g}}
\def\h{\mathfrak{h}}

\def\ad{\text{ad}}

\def\a{\alpha}

\def\pp{\partial_+}
\def\pmin{\partial_-}
\def\pz{\partial_z}
\def\pzb{\partial_{\bar z}}

\def\Dp{D_+}
\def\Dmin{D_-}
\def\Dz{D_z}
\def\Dzb{D_{\bar z}}

\begin{document}
\title[Generalized kappa-deformed spaces]{Generalized kappa-deformed spaces, star-products, and their realizations}
\author{S. Meljanac, S. Kre\v{s}i\'{c}-Juri\'{c}}

\address[S. Meljanac]{Rudjer Bo\v{s}kovi\'{c} Institute, Bijeni\v{c}ka cesta b.b., 10000 Zagreb, Croatia}
\address[S. Kre\v{s}i\'{c}-Juri\'{c}]{Faculty of Natural and Mathematical Sciences, University of Split, Teslina 12, 21000 Split, Croatia}

\email{skresic@fesb.hr}

\maketitle

\begin{abstract}
In this work we investigate generalized kappa-deformed spaces. We develop a systematic method for
constructing realizations of noncommutative (NC) coordinates as formal power series in the Weyl
algebra. All realizations are related by a group of similarity transformations, and to each
realization we associate a unique ordering prescription. Generalized derivatives, the Leibniz rule
and coproduct, as well as the star-product are found in all realizations. The star-product and
Drinfel'd twist operator are given in terms of the coproduct, and the twist operator is derived
explicitly in special realizations. The theory is applied to a Nappi-Witten type of NC space.
\end{abstract}


\section{Introduction}

Recently there has been a growing interest in the formulation of physical theories on
noncommutative (NC) spaces. The structures of such theories and their physical consequences were
studied in \cite{1}-\cite{7}. Classification of NC spaces and investigation of their properties, in
particular the development of a unifying approach to a generalized theory suitable for physical
applications, is an important problem. In order to make a contribution in this direction we analyze
a Lie algebra type NC space which is a generalized version of the kappa-deformed space.

Kappa-deformed spaces were studied by different groups, from both the mathematical and physical
point of view \cite{8}-\cite{32}. There is also an interesting connection between the
kappa-deformed spaces and Doubly Special Relativity program \cite{17}, \cite{18}. In a
kappa-deformed space the noncommutative coordinates satisfy Lie algebra type relations depending on
a deformation parameter $a\in \R^n$. The parameter $a$ is on a very small length scale and yields
the undeformed space when $\|a\|\to 0$. Other types of NC spaces frequently studied in the
literature are the canonical theta-deformed spaces where the corresponding commutation relations
are given by a second rank antisymmetric tensor $\theta_{\mu\nu}$, see \cite{3}, \cite{4} and
references therein.

A simple unification of kappa and theta-deformed spaces was first used in the study of the
Wess-Zumino-Witten model \cite{33}. Unification of these spaces was also the starting point in the
algebraic study of the time-dependent NC geometry of a six dimensional Cahen-Wallach pp-wave string
backgroud \cite{34}-\cite{36}. In this approach the unification is achieved by adding a central
element to the NC coordinates whose commutation relations are parametrized by real-valued
parameters $a$ and $\theta$ which are assumed to be equal.

The motivation for the present work is twofold. First, we want to generalize the unified kappa and
theta-deformed spaces to arbitrary dimensions, and for arbitrary values of the parameters $a$ and
$\theta$. Second, we want to develop a unifying approach to constructing realizations of such
spaces in terms of ordinary commutative coordinates $x_\mu$ and derivatives $\partial_\mu$ which
are convenient for physical applications. In the present work we assume that $x_\mu$ are
coordinates in a Euclidean space, but the analysis can be easily extended to Minkowski space. We
shall be mainly concerned with the Nappi-Witten type of NC space which arises in the study of
pp-wave string background \cite{36}. This space is made up of two copies of kappa-deformed space
and one copy of theta-deformed space. Our analysis is based on the methods developed for algebras
of deformed oscillators and the corresponding creation and annihilation operators
\cite{37}-\cite{46}. The realization of a general Lie algebra type NC space in symmetric Weyl
ordering was given in \cite{47}.

The outline of the paper and the summary of the main results is as follows. In Sec. 2 we introduce
a generalized kappa-deformed space of Nappi-Witten type, $NW_4$. We study realizations of the
generators of $NW_4$ as formal power series with coefficients in the Weyl algebra. We show that
there exist an infinite family of such realizations parameterized by two functions $\varphi$ and
$F$. For special choices of $\varphi$ and $F$ we obtain three important realizations: the right,
symmetric left-right, and Weyl realization. In Sec. 3 we construct a group of similarity
transformations acting transitively on the realizations. Sec. 4 deals with ordering prescriptions
for $NW_4$. We show that to each realization one can associate an ordering prescription, and we
find the prescriptions explicitly in terms of the parameter functions $\varphi$ and $F$. In our
approach the right, symmetric left-right and Weyl realizations correspond to the time, symmetric
time and Weyl orderings defined in \cite{36}, respectively. Thus the orderings found in \cite{36}
are only special cases of an infinite family of ordering prescriptions for $NW_4$ constructed here.
Furthermore, the time and symmetric time orderings can be viewed as limiting cases of an ordering
prescription which interpolates between the two orderings.

In Sec. 5 we consider the problem of extending the NC space $NW_4$ by generalized derivatives such
that the extended space is a deformed Heisenberg algebra. We also define rotation operators on the
extended space which generate the undeformed $so(4)$ algebra. The generalized derivatives and
rotation operators are found in all realizations of $NW_4$. Sec. 6 deals with Leibniz rule and
coproduct for the deformed Heisenberg algebra introduced in Sec. 5. We find explicitly the
coproduct depending on the parameter functions $\varphi$ and $F$, and we give a relation between
the coproducts in different realizations. Furthermore, star-products and Drinfel'd twist operators
are considered in Sec. 7. A general formula for the star-product in terms of the coproduct is
given, and an expression depending on $\varphi$ and $F$ is derived. Also, the corresponding twist
operator is found explicitly for a wide class of realizations of $NW_4$. Finally, we describe how
the obtained results generalize to higher dimensions.

\section{Realizations of the Nappi-Witten space $NW_4$}

Let us consider a unification of the canonical theta-deformed NC space and a Lie algebra type NC
space with generators $X_1,X_2,\ldots ,X_n$ and structure constansts $C_{\mu\nu\lambda}$.
Throughout the paper capital letters will be used consistently to denote NC coordinates. In order
to include the theta-deformation given by a constant anytisymmetric tensor $\theta_{\mu\nu}$, we
introduce a central element $X_0$ such that
\begin{equation}\label{2.32}
[X_\mu,X_\nu]=i\theta_{\mu\nu}X_0 +i\sum_{\lambda}C_{\mu\nu\lambda} X_{\lambda}.
\end{equation}
The NC space defined by the commutation relations \eqref{2.32} is also of Lie algebra type provided all the Jacobi
identitities are satisfied:
\begin{align}
&\sum_{\a}\Big(C_{\mu\nu\a} C_{\a\lambda\rho}+C_{\nu\lambda\a}C_{\a\mu\rho}+C_{\lambda\mu\a}C_{\a\nu\rho}\Big)=0, \\
&\sum_{\a}\Big(C_{\mu\nu\a} \theta_{\a\lambda}+ C_{\nu\lambda\a} \theta_{\a\mu}+ C_{\lambda\mu\a}
\theta_{\a\nu}\Big)=0.
\end{align}
When $\theta_{\mu\nu}\to 0$ we obtain a Lie algebra type NC space with structure constants $C_{\mu\nu\lambda}$. Similarly,
when $C_{\mu\nu\lambda}\to 0$ the space reduces to the canonical theta-deformed space with the additional
central element $X_0$.

As an example consider a NC space with coordinates
$X_+,X_-,Z_\mu,\bar Z_\mu$, $\mu=1,\ldots ,n$, satisfying the commutation relations
\begin{align}
[X_-,Z_\mu] &= -iaZ_\mu, \label{2.33} \\
[X_-,\bar Z_\mu] &= ia\bar Z_\mu,  \label{2.34} \\
[Z_\mu, \bar Z_\nu] &= i2\theta\delta_{\mu\nu}\, X_+. \label{2.35}
\end{align}
Here $X_+$ is the central element and $a,\theta \in \R$. We shall refer to the NC space defined by
\eqref{2.33}-\eqref{2.35} as the generalized Nappi-Witten space $NW_{2n+2}$. In the special case
when $a=\theta$ and $n=2$, this space was recently studied by Halliday and Szabo in \cite{36}.

Without loss of generality we may assume that $n=1$ since all the results are easily extended to
$n>1$. Thus, we shall consider the NC space $NW_4$ generated by $X_+,X_-,Z$ and $\bar Z$ satisfying
\begin{align}
[X_-,Z] &= -ia Z, \label{2.1}\\
[X_-,\bar Z] &= ia \bar Z, \label{2.2} \\
[Z, \bar Z] &= i2\theta X_+.  \label{2.3}
\end{align}
The space $NW_4$ may be considered a generalized kappa-deformed space since \eqref{2.3} defines a
theta-deformation while \eqref{2.1} and \eqref{2.2} define two kappa-deformations. Since $NW_4$ is
a Lie algebra, in future reference it will be denoted $\g$.

For notational ease let $X=(X_+,X_-,Z,\bar Z)$, and let $x=(x_+,x_-,z,\bar z)$ be the ordinary
commutative coordinates with the corresponding derivatives $\partial=(\pp,\pmin,\pz,\pzb)$. We seek
a realization of the generators of $\g$ as formal power series with coefficients in the Weyl
algebra $\mathcal{A}_4$ generated by $x$ and $\partial$. Let us consider realizations of the form
\begin{equation}\label{2.30}
X_{\mu}=\sum_{\a} x_{\a}\, \phi_{\a\mu}(\partial), \quad \phi_{\a\mu}(0)=\delta_{\a\mu},
\end{equation}
that are linear in $x$ and $\phi_{\a\mu}(\partial)$ is a formal power series in $\partial$. We
assume that there exists a dual relation
\begin{equation}\label{2.31}
x_{\mu}=\sum_{\a} X_{\a} \Phi_{\a\mu}(\partial), \quad \Phi_{\a\mu}(0)=\delta_{\a\mu},
\end{equation}
where $\Phi_{\a\mu}(\partial)$ is also a formal power series in $\partial$. A realization
characterized by the functions $\phi_{\mu\nu}$ will be called a $\phi$-realization. The generators
of $\g$ belong to $\overline{\mathcal{A}}_4$, the formal completion of $\mathcal{A}_4$. One may
also consider realizations in which $x_\a$ is placed to the right of $\phi_{\a\mu}(\partial)$, or
any linear combination of the two types. This is convenient when one requires Hermitian
realizations. Indeed, let $\dag \colon \mathcal{A}_4\to \mathcal{A}_4$ be the Hermitian operator
defined by $x_\mu^\dag = x_\mu$, $\partial_\mu^\dag = -\partial_\mu$ and $(x_\mu\partial_\nu)^\dag
=-\partial_\nu x_\mu$. If \eqref{2.30} is a realization, then
\begin{equation}
X_\mu =\frac{1}{2}\sum_\a \Big(x_\a \phi_{\a\mu}(\partial)+\phi_{\a\mu}(-\partial)\,x_\a\Big)
\end{equation}
is a Hermitian realization since $X_\mu^\dag = X_\mu$. Although such realizations are interesing in
their own right, in this paper we shall restrict our attention to realizations of the type
\eqref{2.30}.

Let us assume the Ansatz
\begin{align}
X_+ &= x_+,  \label{2.4} \\
X_- &= x_- +ia\big[\bar z\, \pzb\, \gamma(A)-z\, \pz\, \gamma(-A)\big]+a\theta x_+ \pz \pzb\, \psi(A),  \label{2.5} \\
Z &= z\varphi(-A)+i\theta x_+\, \pzb\, \eta (-A),  \label{2.6}  \\
\bar Z &= \bar z \varphi(A)-i\theta x_+\, \pz\, \eta (A),  \label{2.7}
\end{align}
where $A=ia\pmin$, and the functions $\varphi, \eta, \gamma$ and $\psi$ satisfy the boundary
conditions $\varphi(0)=1$ and $\gamma(0)$, $\psi(0)$ finite. The Ansatz \eqref{2.4}-\eqref{2.7} is
of a fairly general nature leading to a number of interesting realizations discussed below. The
boundary conditions ensure that in the limit $a,\theta\to 0$ the NC coordinates $X_{\mu}$ become
the commutative coordinates $x_{\mu}$.

Let us analyze the realization \eqref{2.4}-\eqref{2.7}. Define $\sigma \colon \g \to \g$ to be the
antilinear map given by $\sigma (X_\pm)=X_\pm$, $\sigma(Z)=\bar Z$, and $\sigma(\bar Z)=Z$ which
also preserves the Lie bracket, $\sigma([X_\mu,X_\nu])=[\sigma(X_\mu),\sigma(X_\nu)]$. Then
$\sigma$ acts as a formal conjugation, and $\sigma^2=id$. The action of $\sigma$ on the generators
of $\mathcal{A}_4$ is defined in the obvious way: $\sigma(x_\pm)=x_\pm$, $\sigma(z)=\bar z$,
$\sigma(\bar z)=z$ and $\sigma (\partial_{\pm})=\partial_{\pm}$, $\sigma (\pz)=\pzb$, $\sigma
(\pzb)=\pz$, and $\sigma(x_\mu\partial_\nu)=\sigma(x_\mu)\sigma(\partial_\nu)$. The condition
$\sigma(X_-)=X_-$ holds if and only if $\psi$ is an even function, as seen from Eq. \eqref{2.5}.
The commutation relations \eqref{2.1}-\eqref{2.3} imply that the functions $\varphi, \eta, \gamma$
and $\psi$ are constrained by the system of equations
\begin{align}
\gamma (A) - \frac{\varphi^\prime (A)}{\varphi (A)} &= 1,  \label{2.8} \\
\varphi (A) \eta(-A)+\varphi(-A)\eta(A) &= 2,  \label{2.9} \\
\eta^\prime (A)-\gamma(-A)\eta(A)-\psi(A)\varphi(A)+\eta(A) &= 0,  \label{2.10}
\end{align}
where the prime denotes the differentiation with respect to $A$. It is convenient to introduce the auxiliary function
\begin{equation}\label{2.11}
F(A)=\varphi(-A)\eta(A)-1.
\end{equation}
Then Eq. \eqref{2.9} implies that $F$ is odd and, furthermore, $F=0$ if and only if
$\psi=0$. For a given choice of $\varphi$ and $F$ one can uniquely determine the remaining
functions $\eta,\gamma$ and $\psi$. Therefore, the Lie algebra $\g$ admits infinitely many
realizations parameterized by $\varphi$ and $F$ satisfying $\varphi(0)=1$ and $F(0)=0$.

Now we turn our attention to special realizations of $\g$: the right realization, symmetric
left-right and Weyl realization. As noted in the introduction to every realization one can
associate an ordering prescription on the universal enveloping algebra $U(\g)$. The aforementioned
realizations correspond to the time ordering, symmetric time ordering and Weyl symmetric ordering
discussed in \cite{36}, respectively.

\subsection{Special realizations}

\subsubsection{Special realization $\gamma=\gamma_0$}

This subsection deals with the realization \eqref{2.4}-\eqref{2.7} when $\gamma$ is a constant function, $\gamma=\gamma_0$,
and $F=0$. For this choice of the parameters Eqs. \eqref{2.8}-\eqref{2.11} imply that
$\varphi (A)=\eta(A)=\exp((\gamma_0-1)A)$ and $\psi(A)=0$. Hence, the $\gamma=\gamma_0$ realization is given by
\begin{align}
X_+ &= x_+, \\
X_- &= x_-+ia\gamma_0\, (\bar z\pzb-z \pz), \\
Z &= (z+i\theta x_+\, \pzb)\, e^{-(\gamma_0-1)A}, \\
\bar Z &= (\bar z-i\theta x_+\, \pz)\, e^{(\gamma_0-1)A}.
\end{align}
Of particular interest are the realizations with $\gamma_0=1$, $\gamma_0=1/2$ and $\gamma_0=0$:

\noindent\textbf{Right realization}: $\gamma_0=1$
\begin{align}
X_+ &= x_+, \label{2.26} \\
X_- &= x_-+ia(\bar z\pzb-z \pz), \label{2.27} \\
Z &= z+i\theta x_+\, \pzb, \label{2.28} \\
\bar Z &= \bar z-i\theta x_+\, \pz. \label{2.29}
\end{align}

\noindent\textbf{Symmetric left-right realization}: $\gamma_0=1/2$
\begin{align}
X_+ &= x_+, \\
X_- &= x_- +\frac{ia}{2}(\bar z\pzb -z\pz), \\
Z &= (z+i\theta x_+\, \pzb)\, e^{\frac{1}{2}A}, \\
\bar Z &= (\bar z-i\theta x_+\, \pz)\, e^{-\frac{1}{2}A}.
\end{align}

\noindent\textbf{Left realization}: $\gamma_0=0$
\begin{align}
X_+ &= x_+, \\
X_- &= x_-, \\
Z &= (z+i\theta x_+\pzb)\, e^A, \\
\bar Z &= (\bar z-i\theta x_+ \pz)\, e^{-A}.
\end{align}
These realizations will be considered later in more detail when we establish a connection between
realizations and ordering prescriptions.

\subsubsection{Weyl Realization}

The Ansatz \eqref{2.4}-\eqref{2.7} also includes the so-called Weyl realization of $\g$ which
corresponds to the symmetric Weyl ordering on $U(\g)$. In this ordering all monomials in the basis
of $U(\g)$ are completely symmetrized over all generators of $\g$.

To this end we recall the following general result proved in \cite{47}. Consider a Lie algebra over
$\C$ with generators $X_1,X_2,\dots ,X_n$ and structure constants $C_{\mu\nu\lambda}$ satisfying
\begin{equation}\label{2.14}
[X_\mu,X_\nu]=i\sum_{\a=1}^n C_{\mu\nu\a} X_\a.
\end{equation}
The Lie algebra \eqref{2.14} can be given a universal realization in terms of the commutative
coordinates $x_\mu$ and derivatives $\partial_\mu$, $1\leq \mu \leq n$, as follows. Let
$B=[B_{\mu\nu}]$ denote the $n\times n$ matrix of differential operators with elements
\begin{equation}\label{2.15}
B_{\mu\nu} = i\sum_{\a=1}^n C_{\a\nu\mu} \partial_\a,
\end{equation}
and let $p(B)=B/(\exp(B)-1)$ be the generating function for the Bernoulli numbers. Then, one can show
that the generators of the Lie algebra \eqref{2.14} admit the realization
\begin{equation}\label{2.16}
X_{\mu}=\sum_{\a=1}^n x_\a\, p(B)_{\a\mu}.
\end{equation}
This is called the Weyl realization since it gives rise to the symmetric Weyl ordering on the
enveloping algebra of \eqref{2.14}.

We shall use the result \eqref{2.16} in order to obtain the Weyl realization of the Lie algebra
$\g$. Recall that the generators of $\g$ are arranged as $X=(X_+,X_-,Z,\bar Z)$; hence the
structure constants $C_{\mu\nu\lambda}$ can be gleamed off from Eqs. \eqref{2.1}-\eqref{2.3}. Then
Eq. \eqref{2.15} yields
\begin{equation}
B=\begin{pmatrix}
0 & 0 & -i2\theta\, \pzb & i2\theta\, \pz \\
0 & 0 & 0 & 0 \\
0 & ia\, \pz & -ia\, \pmin & 0 \\
0 & -ia\, \pzb & 0 & ia \pmin
\end{pmatrix}.
\end{equation}
The explicit form of the matrix $p(B)$ can be found from the identity
\begin{equation}\label{2.19}
p(B)=-\frac{B}{2}+\frac{B}{2}\coth\Big(\frac{B}{2}\Big).
\end{equation}
One can show by induction that $B^{2n}=A^{2n-2}B^2$, $n\geq 1$, where $A=ia\pmin$, and
\begin{equation}\label{2.17}
B^2=\begin{pmatrix} 0 & 4a\theta \pz\, \pzb & i2\theta \pzb\, A & i2\theta \pz\, A \\
0 & 0 & 0 & 0 \\
0 & -ia\pz\, A & A^2 & 0 \\
0 & -ia\pzb\, A & 0 & A^2\end{pmatrix}.
\end{equation}
Expanding Eq. \eqref{2.19} into Taylor series and collecting the terms with even powers of $A$ we obtain
\begin{equation}\label{2.18}
p(B)=1-\frac{B}{2}+h(A) B^2,
\end{equation}
where we have defined
\begin{equation}
h(A)=\frac{1}{A^2}\left(\frac{A}{2}\coth\Big(\frac{A}{2}\Big)-1\right).
\end{equation}
Now Eqs. \eqref{2.17} and \eqref{2.18} yield
\begin{equation}\label{2.20}
p(B)=\begin{pmatrix} 1 & 4a\theta\, \pz\, \pzb\, h(A) & i\theta\, \pzb\, (1+2Ah(A)) & -i\theta\,
\pz\, (1-2Ah(A)) \\
0 & 1 & 0 & 0 \\
0 & -\frac{ia}{2}\, \pz\, (1+2Ah(A)) & 1+ \frac{A}{2}(1+2Ah(A)) & 0 \\
0 & \frac{ia}{2}\, \pzb\, (1-2Ah(A)) & 0 & 1-\frac{A}{2}(1-2Ah(A))\end{pmatrix}.
\end{equation}
Substituting Eq. \eqref{2.20} into Eq. \eqref{2.16} and simplifying, we obtain the Weyl
realization
\begin{align}
X_+ &= x_+,  \label{2.21} \\
X_- &= x_-+ia\left[\bar z \pzb\left(\frac{1}{1-e^A}+\frac{1}{A}\right)-z\pz\left(\frac{1}{1-e^{-A}}-\frac{1}{A}\right)\right] \label{2.22} \\
&+a\theta x_+\, \pz\, \pzb \, \frac{2}{A}\left(\coth\Big(\frac{A}{2}\Big)-\frac{2}{A}\right),  \label{2.23} \\
Z &= z\, \frac{A}{1-e^{-A}}+i\theta x_+\, \pzb\, \left(\frac{2}{1-e^{-A}}-\frac{2}{A}\right),   \label{2.24} \\
\bar Z &= -\bar z\, \frac{A}{1-e^A}-i\theta x_+\, \pz\, \left(\frac{2}{1-e^A}+\frac{2}{A}\right).
\label{2.25}
\end{align}
It is readily seen that the above realization is a special case of the original Ansatz
\eqref{2.4}-\eqref{2.7} where
\begin{align}
\varphi_s(A) &= \frac{A}{e^A-1},  \label{W1} \\
\psi_s(A) &= \frac{2}{A}\left(\coth\Big(\frac{A}{2}\Big)-\frac{2}{A}\right), \\
\gamma_s(A) &= \frac{1}{A}-\frac{1}{e^A-1}, \\
\eta_s(A) &= 2\gamma_s(A), \\
F_s(A) &= \frac{A}{1-\cosh(A)}+\frac{e^A+1}{e^A-1}.  \label{W2}
\end{align}
As required, these functions can be shown to satisfy the compatibility conditions \eqref{2.8}-\eqref{2.10}.

\section{Similarity transformations}

In this section we discuss similarity transformations which connect different realizations
\begin{equation}\label{3.9}
X_{\mu}=\sum_{\a=1}^n x_\a\, \phi_{\a \mu}(\partial), \quad \phi_{\a\mu}(0)=\delta_{\a\mu}.
\end{equation}
The transformations act in a covariant way in the sense that the transformed realization is of the
same type. These transformations can be used to generate new realizations of $\g$ and new ordering
prescriptions on $U(\g)$. They also relate the star-products and coproducts in different
realizations, as discussed in Secs. 6 and 7.

Let $\A_n$ denote the Weyl algebra generated by $x_\mu$ and $\partial_{\mu}$, $1\leq \mu\leq n$.
Consider a differential operator $S$ of the form
\begin{equation}\label{4.08}
S=\exp\Big(\sum_{\a} x_{\a=1}^n A_{\a}(\partial)\Big)
\end{equation}
where $A_{\a}(\partial)$ is a formal power series of $\partial=(\partial_1,\partial_2, \ldots
,\partial_n)$. We assume that $A_\a (0)=0$. Since the commutator of any two elements of the form
$\sum_{\a}x_{\a}A_{\a}(\partial)$ is again of the same form, it follows from the
Baker-Campbell-Hausdorff (BCH) formula that the family of operators $S$ is a group under
multiplication, with identity $S=1$ when $A_{\a}=0$ for all $\a$. To each operator $S$ we associate
a similarity transformation $T_S\colon \overline{\A}_n \to \overline{\A}_n$, $T_S(u)=SuS^{-1}$. The
transformations $T_S$ form a subgroup of the group of inner automorphisms of $\overline{\A}_n$.

Let us examine the action of $T_S$ on the generators of $\A_n$. If we denote $P=\sum_{\a} x_{\a}A_{\a}(\partial)$, then
\begin{equation}\label{4.01}
T_S(x_\mu)=\exp\left(\ad(P)\right) x_{\mu}.
\end{equation}
By induction one can show that
\begin{equation}
\ad^k (P)x_{\mu} = \sum_{\a=1}^n x_{\a}\, R^{(k)}_{\a\mu}(\partial), \quad k\geq 1,
\end{equation}
where the functions $R^{(k)}_{\a\mu}$ are defined recursively by
\begin{align}
R^{(1)}_{\a\mu}(\partial) &= \frac{\partial A_{\a}}{\partial \partial_{\mu}}, \\
R^{(k)}_{\a\mu}(\partial) &= \sum_{\beta=1}^n\left(\frac{\partial A_{\a}}{\partial
\partial_{\beta}}\, R^{(k-1)}_{\beta\mu}+ \frac{\partial R^{(k-1)}_{\a\mu}}{\partial
\partial_{\beta}}\, A_{\beta}\right), \quad k\geq 2.
\end{align}
Hence, we obtain
\begin{equation}\label{4.02}
T_S(x_{\mu}) = \sum_{\a=1}^n x_{\a}\, \Psi_{\a\mu}(\partial)
\end{equation}
where
\begin{equation}
\Psi_{\a\mu}(\partial) = \sum_{k=1}^\infty \frac{1}{k!}\, R^{(k)}_{\a\mu}(\partial)+\delta_{\a\mu}.
\end{equation}
Similarly, the transformation of $\partial_\mu$ yields
\begin{equation}
T_S(\partial_\mu) = \partial_{\mu}+\sum_{k=1}^\infty \frac{(-1)^k}{k!}\, Q^{k-1} (A_\mu (\partial))
\end{equation}
where the operator $Q$ is defined by
\begin{equation}
Q=\sum_{\a=1}^n A_{\a}(\partial)\, \frac{\partial}{\partial \partial_{\a}}.
\end{equation}
We note that the transformation of $\partial_{\mu}$ is given only in terms of $\partial_1,\partial_2,\ldots ,
\partial_n$, which write symbolically as
\begin{equation}\label{4.03}
T_S(\partial_\mu) = \Lambda_{\mu}(\partial).
\end{equation}
The inverse transformations of $x_{\mu}$ and $\partial_{\mu}$ are of the same type,
\begin{equation}\label{4.06}
T_S^{-1}(x_{\mu}) =\sum_{\a=1}^n x_{\a} \widetilde{\Psi}_{\a\mu}(\partial), \quad
T_S^{-1}(\partial_{\mu}) = \widetilde{\Lambda}_{\mu}(\partial).
\end{equation}
The functions $\Psi_{\a\mu}$ and $\Lambda_{\mu}$ are related through the commutation relations for $\partial_\mu$ and $x_\nu$.
Substituting Eqs. \eqref{4.02} and \eqref{4.03} into the commutator $[T_S(\partial_{\mu}),T_S(x_{\nu})]=\delta_{\mu\nu}$ we find
\begin{equation}\label{4.04}
\sum_{\a=1}^n \frac{\partial \Lambda_{\nu}}{\partial \partial_{\a}}\, \Psi_{\a\mu} =
\delta_{\mu\nu}.
\end{equation}
Define the vector $\Lambda = (\Lambda_1,\Lambda_2,\ldots ,\Lambda_n)$ and matrix $\Psi=[\Psi_{\mu\nu}]$. Then Eq. \eqref{4.04}
implies that
\begin{equation}\label{4.05}
\frac{\partial \Lambda}{\partial \partial} = \Psi^{-1}
\end{equation}
where
\begin{equation}
\frac{\partial \Lambda}{\partial \partial} = \Big[\frac{\partial \Lambda_{\nu}}{\partial \partial_{\mu}}\Big]
\end{equation}
is the Jacobian of $\Lambda$.

To prove the covariance of the realization \eqref{3.9} under the action of $T_S$ consider
\begin{equation}
T_S^{-1}(X_{\mu}) = \sum_{\a=1}^n T_S^{-1} (x_{\a})\, \phi_{\a\mu}\big(T_S^{-1}(\partial)\big).
\end{equation}
Using Eq. \eqref{4.06} the above expression becomes
\begin{equation}\label{4.09}
T_S^{-1}(X_{\mu}) = \sum_{\beta=1}^n x_{\beta}\, \widetilde \phi_{\beta\mu}(\partial)
\end{equation}
where
\begin{equation}
\widetilde \phi_{\beta\mu}(\partial) =\sum_{\a=1}^n \widetilde \Psi_{\beta\a}(\partial)\,
\phi_{\a\mu}\big(\widetilde \Lambda (\partial)\big).
\end{equation}
Let us introduce the new variables $y_{\mu}=T_S(x_{\mu})$ and $\partial^y_{\mu}=T_S(\partial_\mu)$
(which also generate the Weyl algebra $\A_n$). Then Eq. \eqref{4.09} yields
\begin{equation}\label{4.10}
X_{\mu}=\sum_{\beta=1}^n y_{\beta}\, \widetilde \phi_{\beta\mu}(\partial^y),
\end{equation}
proving that the realization \eqref{3.9} is covariant under the change of variables $x_{\mu}\mapsto
Sx_{\mu}S^{-1}$ and $\partial_\mu \mapsto S\partial_{\mu}S^{-1}$. Thus, the similarity
transformation $T_S$ maps the $\phi$-realization \eqref{3.9} to $\widetilde \phi$-realization
\eqref{4.10}.

As an example, consider the right realization \eqref{2.26}-\eqref{2.29}. It can be shown that the
operator $S$ mapping the right realization to the general Ansatz \eqref{2.4}-\eqref{2.7}
(parameterized by $\varphi$ and $F$) is given by
\begin{equation}\label{S}
S=\exp\left(z\pz\, \ln\varphi(-A)+\bar z\pzb\, \ln\varphi(A)+i\theta x_+ \pz\pzb\, F(A)
\frac{\ln\varphi(A) \varphi(-A)}{1-\varphi(A)\varphi(-A)}\right).
\end{equation}
Direct calculation yields
\begin{align}
\Lambda_+(\partial) &= \pp+i\theta \pz\, \pzb\, \frac{F(A)}{\varphi(A)\varphi(-A)}, \label{4.11} \\
\Lambda_-(\partial) &= \pmin, \\
\Lambda_z(\partial) &= \pz\, \frac{1}{\varphi(-A)}, \\
\Lambda_{\bar z}(\partial) &= \pzb\, \frac{1}{\varphi(A)}. \label{4.12}
\end{align}
Now, the functions $\Psi_{\mu\nu}(\partial)$ can be calculated from Eq. \eqref{4.05}. The group of
transformations $T_S$ acts transitively since any two realizations are related by
$T_{S_1}T^{-1}_{S_2}$ where $T_{S_i}$ maps the right realization to the
$(\varphi_i,F_i)$-realization.

\section{Generalized orderings}

When considering the NC space $NW_6$ only three ordering prescriptions have been used in \cite{36} for the construction
of the corresponding star-products: time ordering, symmetric time ordering and symmetric Weyl ordering. The time
ordering is defined by
\begin{equation}\label{5.12}
:e^{ikX}:_t = e^{i(k_z Z + k_{\bar z}\bar Z)}\, e^{ik_-X_-}\, e^{ik_+X_+}
\end{equation}
where we have denoted $k=(k_+,k_-,k_z,k_{\bar z})\in \R^4$ and $kX$ is the Euclidean space scalar
product $kX=k_+X_+ + k_-X_- + k_z Z+k_{\bar z} \bar Z$. Since $X_+$ is the central element the
position of $e^{ik_+X_+}$ is irrelevant. Here we consider only the Euclidean space, but the theory
can be easily generalized to spaces with other signatures, e.g. the Minkowski space. The symmetric
time and symmetric Weyl orderings are defined by
\begin{equation}\label{5.13}
:e^{ikX}:_{st} = e^{i\frac{1}{2}k_-X_-}\, e^{i(k_zZ+k_{\bar z}\bar Z)}\, e^{i\frac{1}{2}k_-X_-}\, e^{ik_+X_+}
\end{equation}
and
\begin{equation}\label{5.14}
:e^{ikX}:_s = e^{i(k_+X_+ + k_-X_- +k_zZ + k_{\bar z}\bar Z)},
\end{equation}
respectively. We  note that the orderings are determined by the position of $X_-$ in the monomial basis of $U(\g)$.
For illustration, the monomials of order three (modulo $X_+$) in the time ordering are
\begin{equation}\label{5.06}
\begin{split}
&X_-^3, Z^3, \bar Z^3, Z^2 X_-, \bar Z^2 X_-, Z X_-^2, \bar Z X_-^2, \frac{1}{2}(Z \bar Z+\bar Z
Z)X_-, \\
&\frac{1}{3}(Z^2 \bar Z+Z\bar Z Z+\bar Z Z^2), \frac{1}{3}(Z \bar Z^2+\bar Z Z \bar Z+\bar Z^2 Z).
\end{split}
\end{equation}
The corresponding monomials in the symmetric time ordering and symmetric Weyl ordering are given by
\begin{equation}\label{5.07}
\begin{split}
&X_-^3, Z^3, \bar Z^3, \frac{1}{2}(Z^2 X_- + X_- Z^2), \frac{1}{2}(\bar Z^2 X_- + X_- \bar Z^2),
\frac{1}{2}(Z X_-^2 + X_-^2 Z), \\
&\frac{1}{2}(\bar Z X_-^2 + X_- \bar Z), \frac{1}{3}(Z^2 \bar Z+ Z\bar Z Z+\bar Z Z^2),
\frac{1}{3}(Z \bar Z^2 + \bar Z Z \bar Z+\bar Z^2 Z), \\
&\frac{1}{4}(Z \bar Z X_- + X_- Z \bar Z + \bar Z Z X_- + X_- \bar Z Z),
\end{split}
\end{equation}
and
\begin{equation}\label{5.08}
\begin{split}
&X_-^3, Z^3, \bar Z^3, \frac{1}{3}(X_-^2 Z+X_- Z X_- + Z X_-^2), \frac{1}{3}(X_-^2 \bar Z + X_-
\bar Z X_- + \bar Z X_-^2), \\
&\frac{1}{3}(Z^2 \bar Z + Z \bar Z Z+\bar Z Z^2), \frac{1}{3}(\bar Z^2 Z + \bar Z Z \bar Z + Z \bar
Z^2), \frac{1}{3}(X_- Z^2 + Z X_- Z + Z^2 X_-), \\
&\frac{1}{3}(X_- \bar Z^2 + \bar Z X_- \bar Z + \bar Z X_-), \frac{1}{6}(X_- Z \bar Z +
\textit{cyclic perm.}),
\end{split}
\end{equation}
respectively. In future reference the time ordering and symmetric time ordering will be called the right
ordering and symmetric left-right ordering, respectively, as implied by the position of $X_-$ in the monomial
basis.

In this section we show that to each realization \eqref{2.4}-\eqref{2.7} of the generators of $\g$
one can associate an ordering prescription on $U(\g)$. This leads to an infinite family of ordering
prescriptions parameterized by the functions $\varphi$ and $F$. In our approach the orderings used
in \cite{36} appear as special cases corresponding to the right, symmetric left-right and Weyl
realization found in Sec. 2.

Let us begin by defining the ``vacuum'' state
\begin{equation}
|0\rangle =1, \quad \partial_{\mu} |0\rangle =0.
\end{equation}
Let $X_\mu^\phi$ denote the generator $X_\mu$ in $\phi$-realization \eqref{3.9}, and let $X_\mu^s$
denote the Weyl realization of $X_\mu$. It has been shown in \cite{25} and \cite{26} that for
kappa-deformed spaces a simple relation holds,
\begin{equation}\label{5.01}
e^{ikX^s} |0\rangle = e^{ikx}, \quad k\in \R^n.
\end{equation}
Eq. \eqref{5.01} can be generalized to any $\phi$-realization by requiring
\begin{equation}
e^{iK(k)X^\phi}|0\rangle = e^{ikx}
\end{equation}
for some function $K\colon \R^n\to \R^n$. Let $T_S$ be the similarity transformation mapping the
$\phi$-realization \eqref{3.9} to the symmetric Weyl realization
\begin{equation}
X_{\mu}^s = \sum_\a y_\a\, \phi_{\a\mu}^s (\partial^y),
\end{equation}
where the variables $y_\mu$ and $\partial^y_\mu$ are given by Eqs. \eqref{4.02} and \eqref{4.03}.
In this realization we have
\begin{equation}
e^{iK(k)X^s}|0\rangle = e^{iK(k)y},
\end{equation}
hence $e^{iK(k)y}=e^{ikx}$. Since $\partial^y_\mu=\Lambda_\mu(\partial)$, it follows that
$\partial^y_\mu\, e^{iK(k)y}=\Lambda_\mu(\partial) e^{ikx}$ which implies
\begin{equation}\label{5.09}
K_{\mu}(k)=-i\Lambda_{\mu}(ik), \quad 1\leq \mu \leq n.
\end{equation}
Thus, $K_\mu$ is completely determined by the similarity transformation $T_S$ mapping the $\phi$-realization to the Weyl realization.

For each $\phi$-realization we define the $\phi$-ordering by
\begin{equation}\label{5.03}
: e^{ikX}:_{\phi} = e^{iK(k)X}
\end{equation}
where $K$ is given by Eq. \eqref{5.09}. If $X$ is represented in $\phi$-realization, then
\begin{equation}\label{5.15}
: e^{ikX}:_{\phi} |0\rangle = e^{ikx}.
\end{equation}
The above expression gives a simple relation between a $\phi$-realization and $\phi$-ordering.
The monomial basis for $U(\g)$ in $\phi$-ordering can be explicitly derived from Eq. \eqref{5.03}.
Let $m=(m_i)$ be a multi-index with $m_i\in \mathbb{N}_0$, and let
\begin{equation}
\left(\frac{\partial}{\partial k}\right)^m = \frac{\partial^{|m|}}{\partial k_1^{m_1}\ldots \partial k_n^{m_n}}.
\end{equation}
A basis element of order $|m|$ is given by
\begin{equation}\label{5.04}
P_{m}(X)=\left(-i\frac{\partial}{\partial k}\right)^m e^{iK(k)X}\Big|_{k=0}.
\end{equation}
Since $K(0)=0$, $P_{m}(X)$ is a polynomial of degree $|m|$. In the Weyl realization when $K(k)=k$, Eq. \eqref{5.04}
leads to the Weyl ordering whereby the polynomials $P_{m}(X)$ are completely symmetrized over the generators of $\g$.

Let us illustrate the above ideas by computing an ordering prescription for the NC space $NW_4$.
For a general realization parameterized by $\varphi$ and $F$ we have
\begin{equation}\label{K0}
 :e^{ikX}:_{(\varphi,F)}=e^{i(K_+(k)X_+ +K_-(k)X_- + K_z(k)Z+K_{\bar z}(k)\bar Z)}.
\end{equation}
One can use Eqs. \eqref{4.11}-\eqref{4.12} to find the similarity transformation mapping the
$(\varphi,F)$-realization to Weyl realization. Then, Eq. \eqref{5.09} yields
\begin{align}
K_+(k) &= k_+ -\theta k_z k_{\bar z}\, \frac{F(-ak_-)-F_s(-ak_-)}{\varphi(-ak_-) \varphi(ak_-)},  \label{K1} \\
K_-(k) &= k_-, \label{K2} \\
K_z(k) &=k_z\, \frac{\varphi_s(ak_-)}{\varphi(ak_-)}, \label{K3} \\
K_{\bar z}(k) &= k_{\bar z}\, \frac{\varphi_s(-ak_-)}{\varphi(-ak_-)},  \label{K4}
\end{align}
where $\varphi_s$ and $F_s$ are the parameter functions defined by Eqs. \eqref{W1} and \eqref{W2}.
Thus, Eqs. \eqref{K0}-\eqref{K4} define an infinite family of orderings on $U(\g)$ depending on the
parameter functions $\phi$ and $F$.

Of particular interest is the realization $\gamma=\gamma_0$, in which case
\begin{equation}
\varphi(ak_-)=\exp\big((\gamma_0-1)ak_-\big), \quad F(ak_-)=0.
\end{equation}
In this realization the function $K$ becomes
\begin{equation}
K(k)=\left(k_+ +\theta k_zk_{\bar z} F_s(-ak_-), k_-, k_z\, \frac{\varphi_s(ak_-)}{\varphi(ak_-)}, k_{\bar z}\,
\frac{\varphi_s(-ak_-)}{\varphi(-ak_-)}\right).
\end{equation}
It can be shown that the ordering induced by this realization can be written in exponential form as
\begin{equation}\label{5.11}
e^{i(1-\gamma_0)k_- X_-}\, e^{i(k_z Z+k_{\bar z}\bar Z)}\, e^{i\gamma_0 k_- X_-}=
e^{i\left(k_- X_- + k_z\, \frac{\varphi_s(ak_-)}{\varphi(ak_-)}Z + k_{\bar z}\,
\frac{\varphi_s(-ak_-)}{\varphi(-ak_-)}\bar Z\right)}
\end{equation}
where the central element $X_+$ has been left out. The above ordering has three interesting cases:
the right ordering for $\gamma_0=1$, left ordering for $\gamma_0=0$ and symmetric left-right
ordering for $\gamma_0=1/2$. Therefore, Eq. \eqref{5.11} may be interpreted as an interpolation
between the left and right ordering. A comparison with Eqs. \eqref{5.12}-\eqref{5.13} shows that
the right and symmetric left-right orderings are precisely the time and symmetric time orderings
constructed in \cite{36}.

\section{Generalized derivatives}

This section is devoted to extensions of the Lie algebra $\g$ defined by \eqref{2.1}-\eqref{2.3} by
addition of generalized derivatives. The motivation for considering such extensions is to extend
the deformation of the commutative space to the entire phase space. For a general Lie algebra type
NC space a detailed treatment of the generalized derivatives may be found in \cite{48}. Here we
consider extensions of $\g$ such that the generalized derivatives and the Lie algebra $\g$ are
complementary subalgebras of a deformed Heisenberg algebra $\h$. A natural way to define $\h$ is as
follows. If the generators of $\g$ are given by $\phi$-realization \eqref{2.30}, we define the
generalized derivative $D_\mu$ by setting $D_\mu = \partial_\mu$. Then the commutation relations
yield $[D_\mu,D_\nu]=0$ and $[D_\mu,X_\nu]=\phi_{\mu\nu}(D)$. Furthermore, the Jacobi identitites
are satisfied for all combinations of the generators $X_\mu$ and $D_\nu$. In the limit as $a,\theta
\to 0$ we have $[D_\mu,X_\nu]=\delta_{\mu\nu}$, hence $\h$ is a deformed Heisenberg algebra.
Obviously, there are infinitely many such extensions depending on the realization $\phi$. In our
case the simplest extension is obtained in the right realization \eqref{2.26}-\eqref{2.29} when
$\phi_{\mu\nu}$ are linear in $D_\mu$. Then $\h$ is a Lie algebra defined by the commutation
relations \eqref{2.1}-\eqref{2.3} and
\begin{alignat}{2}
[\Dp,X_+] &= 1, & \qquad [\Dp,Z] &= i\theta \Dzb,  \label{H.1} \\
[\Dp,X_-] &= 0, & \qquad [\Dp,\bar Z] &= -i\theta \Dz, \label{H.2} \\
[\Dmin,X_+] &= 0, & \qquad [\Dmin,Z] &= 0, \label{H.3} \\
[\Dmin,X_-] &= 1, & \qquad [\Dmin,\bar Z] &= 0, \label{H.4} \\
[\Dz,X_+] &= 0, & \qquad [\Dz,Z] &= 1,  \label{H.5} \\
[\Dz,X_-] &= -ia \Dz, & \qquad [\Dz,\bar Z] &= 0, \label{H.6} \\
[\Dzb,X_+] &= 0, & \qquad [\Dzb,Z] &= 0,  \label{H.7} \\
[\Dzb,X_-] &= ia \Dzb, & \qquad [\Dzb,\bar Z] &= 1. \label{H.8}
\end{alignat}
This algebra agrees with the deformed Heisenberg algebra discussed in \cite{36}. We note that if
any other realization is used for the construction of $\h$, then $\h$ will not be of Lie type since
$\phi_{\mu\nu}(D)$ is generally a formal power series in $D_\mu$.

Now let us fix the algebra $\h$ as above. We seek realizations of $\h$ when the NC coordinates
$X_\mu$ are given by the general Ansatz \eqref{2.4}-\eqref{2.7}. Accordingly, we no loger have
$D_\mu=\partial_\mu$ since the realizations of $D_\mu$ must be modified in order to satisfy the
commutators \eqref{H.1}-\eqref{H.8}. First, we consider the derivatives $D_z$ and $D_{\bar z}$.
Assume that $\Dz=\pz\, G(A)$ where $G(A)$ is a formal power series in $A$ satisfying the boundary
condition $G(0)=1$. The boundary condition ensures that $\Dz\to \pz$ as $a\to 0$. Using Eqs.
\eqref{2.5} and \eqref{2.8} one can show that $[\Dz, X_-]=-ia \Dz$ if and only if
\begin{equation}\label{3.3}
\frac{G^\prime (A)}{G(A)} = \frac{\varphi^\prime (-A)}{\varphi (-A)}.
\end{equation}
The unique solution of this equation satisfying $G(0)=1$ is given by $G(A)=1/\varphi (-A)$. Hence,
\begin{equation}\label{3.4}
\Dz = \pz\, \frac{1}{\varphi (-A)}.
\end{equation}
One can verify that if $D_z$ is given by Eq. \eqref{3.4}, then the remaining commutators with $\Dz$ are automatically satisfied.
A similar argument yields
\begin{equation}\label{3.5}
\Dzb = \pzb\, \frac{1}{\varphi (A)}.
\end{equation}
Next we consider $\Dp$. The relation $[\Dp,X_+]=1$ implies that
\begin{equation}\label{3.6}
\Dp=\pp+H(\pmin,\pz,\pzb)
\end{equation}
for some function $H$. Substituting Eqs. \eqref{2.5} and \eqref{3.6} into the commutator
$[\Dp,X_-]=0$ we obtain
\begin{equation}\label{3.7}
a\theta \pz \pzb\, \psi(A)+\frac{\partial H}{\partial \pmin}+ia \frac{\partial H}{\partial \pzb}\, \pzb\, \gamma(A)
-ia \frac{\partial H}{\partial \pz}\, \pz\, \gamma(-A) = 0.
\end{equation}
Similarly, the commutator $[\Dp,Z]=i\theta \Dzb$ yields
\begin{equation}\label{3.8}
i\theta \pzb\, \eta(-A) + \frac{\partial H}{\partial \pz} \varphi(-A)=i\theta \pzb\, \frac{1}{\varphi (A)}.
\end{equation}
The structure of Eq. \eqref{3.7} suggests that $H$ is of the form $H=\pz\, \pzb\, H_0(A)$.
Inserting this expression into Eq. \eqref{3.8} we obtain
\begin{equation}
i\theta \eta(-A) \varphi (A) + H_0(A) \varphi(A) \varphi (-A) = i\theta.  \label{3.10}
\end{equation}
Eqs. \eqref{3.10} and \eqref{2.11} imply that $H_0(A)\varphi(A)\varphi(-A)=i\theta F(A)$, which
yields
\begin{equation}\label{3.11}
H=i\theta \pz\, \pzb\, \frac{F(A)}{\varphi (A)\varphi (-A)}.
\end{equation}
One can verify that the above expression for $H$ is consistent with Eq. \eqref{3.7}. Therefore,
$\Dp$ is given by
\begin{equation}\label{3.12}
\Dp =\pp +i\theta \pz\, \pzb\, \frac{F(A)}{\varphi (A) \varphi (-A)}.
\end{equation}
As required, the remaining commutator $[\Dp, \bar Z]=-i\theta \Dz$ is automatically satisfied.
Finally, we observe that the relations \eqref{H.3} and \eqref{H.4} trivially hold if we define $\Dmin=\pmin$, hence
$\Dmin$  is the same in all realizations.

For future reference we collect the realizations of $D_\mu$ obtained here:
\begin{align}
\Dp &= \pp + i\theta \pz\, \pzb\, \frac{F(A)}{\varphi (A) \varphi (-A)},  \label{D.1} \\
\Dmin &= \pmin, \label{D.2} \\
\Dz &= \pz\, \frac{1}{\varphi (-A)},  \label{D.3} \\
\Dzb &= \pzb\, \frac{1}{\varphi (A)}. \label{D.4}
\end{align}
Comparing the expressions for $D_\mu$ with Eqs. \eqref{4.11}-\eqref{4.12} we conclude that the
generalized derivatives are given by the similarity transformation $D_\mu =S\partial_\mu S^{-1}$
where $S$ is defined by Eq. \eqref{S}. We note that for $\varphi(A)=1$ and $F(A)=0$ (right
realization) we obtain $D_\mu=\partial_\mu$, as required.

The Lie algebra $\h$ can be extended further by adding the rotation generators $M_{\mu\nu}$ which form the ordinary
rotation algebra $so(4)$. The rotation generators are defined by $M_{\mu\nu} = x_\mu \partial_\nu - x_\nu \partial_\mu$,
and satisfy
\begin{align}
M_{\mu\nu} &= -M_{\nu\mu}, \\
[M_{\mu\nu},M_{\lambda\rho}] &= \delta_{\nu\lambda} M_{\mu\rho}-\delta_{\mu\lambda} M_{\nu\rho}-\delta_{\nu\rho} M_{\mu\lambda}
+\delta_{\mu\rho} M_{\nu\lambda}.
\end{align}
Suppose that $X_\mu$ is represented in the right realization \eqref{2.26}-\eqref{2.29}, and let $D_\mu = \partial_\mu$.
By solving Eqs. \eqref{2.26}-\eqref{2.29} for $x_\mu$ the rotation generators can be expressed in terms of $X_\mu$ and $D_\nu$ as
\begin{align}
M_{+-} &= X_+ D_- - X_- D_+ - ia(Z D_z-\bar Z D_{\bar z})D_+ - 2a\theta X_+ D_+ D_z D_{\bar z}, \\
M_{+z} &= X_+ D_z - Z D_+ +i\theta X_+ D_+ D_{\bar z}, \\
M_{+\bar z} &= X_+ D_{\bar z} - \bar Z D_+ -i\theta X_+ D_+ D_z, \\
M_{-z} &= X_- D_z - Z D_- +ia(Z D_z - \bar Z D_{\bar z})D_z + i\theta X_+ D_- D_{\bar z} + 2a\theta X_+ D^2_z D_{\bar z}, \\
M_{-\bar z} &= X_- D_{\bar z} - \bar Z D_- -ia(\bar Z D_{\bar z} -Z D_z)D_{\bar z} -i\theta X_+ D_- D_z +2a\theta X_+ D_z D^2_{\bar z}, \\
M_{z\bar z} &= Z D_{\bar z} - \bar Z D_z -i\theta X_+ (D_z^2+D^2_{\bar z}).
\end{align}
Now the commutators $[M_{\mu\nu},X_\lambda]$ and $[M_{\mu\nu},D_\lambda]$ can be easily found,
which we omit here. In this case the commutators $[M_{\mu\nu},X_\lambda]$ are not linear in
$M_{\mu\nu}$ and $X_\lambda$. We note, however, that by choosing a different realization of $X_\mu$
the rotation generators may be constructed so that the commutators $[M_{\mu\nu},X_\lambda]$ are of
Lie algebra type. For a different approach to construction of $M_{\mu\nu}$ in kappa-deformed spaces
see \cite{9}, \cite{10}, \cite{25} and \cite{26}.

\section{The Leibniz rule and coproduct}

Having introduced generalized derivatives we now set to find the Leibniz rule for $D_\mu$ and the
corresponding coproduct $\Delta D_\mu$. The generalized derivative $D_\mu$ induces a linear map
$D_\mu\colon U(\g)\to U(\g)$ defined as follows. For $1,X_\mu \in U(\g)$ we define $D_\mu \bullet 1
= 0$ and $D_\mu \bullet X_\nu = [D_\mu,X_\nu]\bullet 1$, where $[D_\mu,X_\nu]\bullet 1$ is
calculated using the commutation relations \eqref{H.1}-\eqref{H.8}. The action of $D_\mu$ on
monomials of higher degree can be defined inductively. Suppose we have defined $D_\mu \bullet f(X)$
where $f(X)\in U(\g)$ is a monomial of order $n$. Then the action of $D_\mu$ on a monomial of order
$n+1$ is defined by
\begin{equation}
D_\mu\bullet (X_\nu f(X)) = X_\nu D_\mu \bullet f(X)+[D_\mu,X_\nu]\bullet f(X).
\end{equation}
Since $[D_\mu,X_\nu]$ is given by Eqs. \eqref{H.1}-\eqref{H.8}, by induction hypothesis $[D_\mu,X_\nu]\bullet f(X)$
is well-defined.

One can show that the Leibniz rule for $D_\mu$ is given by
\begin{align}
D_+\bullet (f(X)g(X)) &= (D_+\bullet f(X))g(X)+f(X)\left(D_+\bullet g(X)\right) \notag \\
&+i\theta e^A\, \left(D_z\bullet f(X)\right) \left(D_{\bar z}\bullet g(X)\right) \notag \\
&-i\theta e^{-A}\, \left(D_{\bar z}\bullet f(X)\right)
\left(D_z\bullet g(X)\right), \\
D_-\bullet (f(X)g(X)) &= \left(D_-\bullet f(X)\right) g(X)+f(X)\left(D_-\bullet g(X)\right), \\
D_z\bullet (f(X)g(X)) &= \left(D_z \bullet f(X)\right) g(X)+e^{-A} f(X)\left(D_z\bullet g(X)\right), \\
D_{\bar z}\bullet (f(X)g(X)) &= \left(D_{\bar z}\bullet f(X)\right) g(X)+e^A f(X)\left(D_{\bar z}\bullet g(X)\right),
\end{align}
where $A=iaD_-$. From the above relations one obtains the corresponding coproduct
\begin{align}
\Delta D_+ &= D_+\otimes 1 + 1\otimes D_+ +i\theta e^A\, D_z\otimes D_{\bar z}-i\theta e^{-A}\, D_{\bar z}\otimes D_z, \label{CD.1}\\
\Delta D_- &= D_-\otimes 1+1\otimes D_-, \label{CD.2} \\
\Delta D_z &= D_z\otimes 1+e^{-A}\otimes D_z, \label{CD.3} \\
\Delta D_{\bar z} &= D_{\bar z}\otimes 1+e^A\otimes D_{\bar z}. \label{CD.4}
\end{align}
If $a=\theta$, this coproduct agrees with the time-ordered coproduct found in \cite{36}.

Now let us consider the algebra generated by $X_\mu$ and $\partial_\mu$ where $X_\mu$ is given by
the Ansatz \eqref{2.4}-\eqref{2.7}. The generators satisfy the commutation rules
\begin{alignat}{2}
[\pp,X_+] &= 1,                              & \qquad [\pp,Z] &= i\theta \pzb\, \eta(-A), \label{L.1} \\
[\pp,X_-] &= a\theta \pz\, \pzb\, \psi(A), & \qquad [\pp,\bar Z] &= -i\theta \pz\, \eta(A), \label{L.2} \\
[\pmin,X_+] &=0,                             & \qquad [\pmin,Z] &= 0, \label{L.3} \\
[\pmin,X_-] &=1,                             & \qquad [\pmin,\bar Z] &= 0, \label{L.4} \\
[\pz,X_+] &= 0,                              & \qquad [\pz,Z] &= \varphi(-A) , \label{L.5} \\
[\pz,X_-] &= -ia\pz\, \gamma(-A),            & \qquad [\pz,\bar Z] &= 0, \label{L.6} \\
[\pzb,X_+] &= 0,                             & \qquad [\pzb,Z] &=0, \label{L.7} \\
[\pzb,X_-] &=ia\pzb\, \gamma(A),             & \qquad [\pzb,\bar Z] &= \varphi(A).  \label{L.8}
\end{alignat}
This is also a deformed Heisenberg algebra, as seen by taking the limit $a,\theta \to 0$. The
coproduct $\Delta \partial_\mu$ can be found from the coproduct of $D_\mu$ and Eqs.
\eqref{D.1}-\eqref{D.4} relating $D_\mu$ and $\partial_\mu$. From Eq. \eqref{D.2} we have
\begin{equation}
\Delta \pmin = \pmin \otimes 1 + 1\otimes \pmin,
\end{equation}
which implies that $\Delta A=A\otimes 1 + 1\otimes A$. For convenience let us denote $A_1=A\otimes 1$
and $A_2=1\otimes A$, so that $\Delta A=A_1+A_2$. Furthermore, using Eqs. \eqref{D.3} and
\eqref{CD.3} we find
\begin{align}\label{Dpz}
\Delta \pz &= \Delta D_z \Delta \varphi(-A) \notag \\
&= \varphi(-A_1-A_2)\left(\frac{\pz}{\varphi(-A)}\otimes 1 + e^{-A}\otimes \frac{\pz}{\varphi(-A)}\right),
\end{align}
where we have used commutativity of $D_z$ and $\varphi(A)$. Similarly, one obtains
\begin{equation}\label{Dpzb}
\Delta \pzb = \varphi(A_1+A_2)\left(\frac{\pzb}{\varphi(A)}\otimes 1+e^A \otimes \frac{\pzb}{\varphi(A)}\right).
\end{equation}
It follows from Eq. \eqref{D.1} that
\begin{equation}
\Delta \pp = \Delta D_+ -i\theta \Delta \pz\, \Delta \pzb\, \frac{F(A_1+A_2)}{\varphi(A_1+A_2)\varphi(-A_1-A_2)}.
\end{equation}
Inserting Eqs. \eqref{Dpz} and \eqref{Dpzb} into the above expression we obtain
\begin{align}
\Delta \pp &= \Delta D_+  \notag \\
&-i\theta \left(\frac{\pz}{\varphi(-A)}\otimes 1+e^{-A}\otimes \frac{\pz}{\varphi(-A)}\right)
\left(\frac{\pzb}{\varphi(A)}\otimes 1+e^A \otimes \frac{\pzb}{\varphi(A)}\right) F(A_1+A_2).
\end{align}
If the coproduct $\Delta D_+$ in Eq. \eqref{CD.1} is expressed in terms of $\partial_\mu$ using Eqs. \eqref{D.1}-\eqref{D.4},
then after simplifying one can show that $\Delta \pp$ takes the form
\begin{align}
\Delta \pp &= \pp\otimes 1+1\otimes \pp+i\theta \frac{\pz\, \pzb}{\varphi(A)\varphi(-A)}\otimes 1
\Big(F(A)\otimes 1-F(A_1+A_2)\Big) \notag \\
&+i\theta\, 1\otimes \frac{\pz\, \pzb}{\varphi(A)\varphi(-A)}\Big(1\otimes F(A)-F(A_1+A_2)\Big) \notag  \\
&+i\theta e^A\, \frac{\pz}{\varphi(-A)}\otimes \frac{\pzb}{\varphi(A)}\Big(1\otimes 1-F(A_1+A_2)\Big) \notag \\
&-i\theta e^{-A}\, \frac{\pzb}{\varphi(A)}\otimes \frac{\pz}{\varphi(-A)}\Big(1\otimes 1+F(A_1+A_2)\Big).
\end{align}
The coproduct $\Delta \partial_\mu$ is fixed by the realization $(\varphi,F)$. If $\varphi$ and $F$
parametrize the Weyl realization (c.f. Eqs. \eqref{W1}-\eqref{W2}) and $a=\theta$, then $\Delta
\partial_\mu$ yields the Weyl ordered coproduct found in \cite{36}. Recall that all realizations are
related by similarity transformations described in Sec. 3. Thus, if the coproduct is known in one
realization, then it is known in any other realization. Hence, $\Delta \partial_\mu$ is unique in
the sense that there is only one equivalence class $[\Delta \partial_\mu]$ containing all the
coproducts found above.

\section{Star products and twists}

In this section we study isomorphisms between the spaces of smooth functions of commutative
coordinates $x_\mu$ and NC coordinates $X_\mu$. These isomorphisms are defined in terms of
$\phi$-realizations of $X_\mu$ given by Eq. \eqref{2.30}.

We define the $\phi$-induced isomorphism $\Omega_{\phi}$ by
\begin{equation}
\Omega_\phi f(x) = f(x)|0\rangle \equiv \hat f_\phi (X).
\end{equation}
Here $f(x)|0\rangle$ is calculated by expressing $x_\mu$ in terms of $X_\mu$ and $\partial_\mu$
from Eq. \eqref{2.31}, and placing $\partial_\mu$ to the far right using the commutation relations
$[\partial_\mu, X_\nu] = \phi_{\mu\nu}(\partial)$. Similarly, the inverse map is defined by
\begin{equation}
\Omega_\phi^{-1} \hat f(X) = \hat f(X)|0\rangle \equiv f_\phi (x),
\end{equation}
where $\hat f(X)|0\rangle$ is calculated using Eq. \eqref{2.30}.

We define the $\phi$-star product of functions $f(x)$ and $g(x)$ by
\begin{equation}
(f\star_\phi g)(x)=\hat f_\phi (X)\, \hat g_\phi(X)|0\rangle.
\end{equation}
The star-product can be written in terms of the isomorphism $\Omega_\phi$ as
\begin{equation}
(f\star_\phi g)(x)=\left(\Omega_\phi f(x)\right) g(x),
\end{equation}
where the derivatives $\partial_\mu$ in $\Omega_\phi (f(x))$ are placed to the far right and act on the
function $g(x)$. The star-product may also be written in the form
\begin{equation}
(f\star_\phi g)= m_0 \mathcal{F}_\phi (f\otimes g)
\end{equation}
where $m_0$ denotes the ordinary pointwise multiplication and $\mathcal{F}_\phi$ is the
corresponding Drinfel'd twist operator \cite{36}. We introduce the $\phi$-deformed multiplication
$m_\phi=m_0 \mathcal{F}$ so that $f\star_\phi g=m_\phi (f\otimes g)$. The isomorphism $\Omega_\phi$
can be written in terms of the twist operator $\mathcal{F}_\phi$ as
\begin{equation}
\left(\Omega_\phi f\right)g = m_0 \mathcal{F}_\phi \left(f\otimes g\right) \quad \forall g.
\end{equation}

Let us now consider the following problem. Given the exponential functions $e^{ikx}$ and $e^{iqx}$,
$k,q\in \R^n$, we want to calculate their star-product in $\phi$-realization. For NC coordinates
$X_\mu$ we have
\begin{equation}\label{6.02}
e^{ikX} e^{iqX}=e^{iD_s(k,q)X}
\end{equation}
where the function $D_s\colon \R^n\times \R^n \to \R^n$ can be found in principle from the Dynkin
form of the BCH formula. If $X$ is represented in the symmetric Weyl realization, denoted $X^s$,
then $e^{ikX^s}|0\rangle = e^{ikx}$. This implies that
\begin{equation}
e^{ikx}\star_s e^{iqx} = e^{iD_s(k,q)x},
\end{equation}
where $\star_s$ denotes the Weyl-ordered star product. The above relations can be generalized to
arbitrary $\phi$-ordering:
\begin{align}
:e^{ikX}:_\phi\, :e^{iqX}:_\phi &=\, :e^{iD_\phi(k,q)X}:_\phi, \\
e^{ikx}\star_\phi e^{iqx} &= e^{iD_\phi (k,q)x}, \label{6.07}
\end{align}
for some function $D_\phi \colon \R^n\times \R^n \to \R^n$. Let us find the correspondence between
$D_\phi$ and $D_s$. Recall from Eq. \eqref{5.03} that in $\phi$-ordering we have
\begin{equation}\label{6.03}
:e^{ikX}:_\phi\, = e^{iK_\phi (k)X}
\end{equation}
where $K_\phi$ is given by the similarity transformation $\Lambda$ which maps the $\phi$-ordering
to symmetric Weyl ordering. Furthermore, it follows from Eq. \eqref{6.02} that
\begin{equation}\label{6.04}
:e^{ikX}:_\phi \, :e^{iqX}:_\phi\, = e^{iD_s(K_\phi(k),K_\phi(q))X}.
\end{equation}
In view of Eq. \eqref{6.03} we have
\begin{equation}
e^{iD_s(K_\phi(k),K_\phi(q))X}=\, :e^{iK^{-1}_\phi \left(D_s(K_\phi(k),K_\phi(q))\right)X}:_\phi,
\end{equation}
hence
\begin{equation}\label{6.06}
:e^{ikX}:_\phi\, :e^{iqX}:_\phi =\, :e^{iK^{-1}_\phi \left(D_s(K_\phi(k),K_\phi(q))\right)X}:_\phi.
\end{equation}
Therefore, the function $D_\phi$ is given by
\begin{equation}
D_\phi(k,q)=K^{-1}_\phi \big(D_s(K_\phi(k),K_\phi(q))\big).
\end{equation}
If the isomorphism $\Omega_\phi$ is restricted to the space of Schwartz functions on $\R^n$, then
for a Schwartz function $f(x)$ we may define the Fourier transform
\begin{equation}
\widetilde f(k) = \frac{1}{(2\pi)^{n/2}}\int_{\R^n} f(x)\, e^{-ikx}\, dx.
\end{equation}
In this case, $\hat f_\phi (X) = \Omega_\phi f(x)$ has the Fourier representation
\begin{equation}\label{6.05}
\hat f_{\phi}(X)=\frac{1}{(2\pi)^{n/2}}\int_{\R^n} \widetilde f(k)\, :e^{ikX}:_\phi\, dk
\end{equation}
defined in terms of the $\phi$-ordering $:e^{ikX}:_\phi$. Using the Fourier representation
\eqref{6.05} and Eq. \eqref{6.07} it can be shown that the general $\phi$-ordered star product can
be expressed in terms of a bi-differential operator as
\begin{equation}
(f\star_\phi g)(x)=e^{ix\left[D_\phi(-i\partial^u,-i\partial^v)+i\partial_u+i\partial^v\right]} f(u)g(v)\Big|_{\substack{u=x\\ v=x}}.
\end{equation}

Next we want to relate the coproduct $\Delta_\phi$ in $\phi$-realization with the function
$D_\phi(k,q)$. Let us start with the undeformed coproduct $\Delta_0$ satisfying
\begin{equation}\label{6.09}
\partial_\mu m_0 = m_0 \Delta_0 \partial_\mu,
\end{equation}
which gives a simple relation between the Leibniz rule for $\partial_\mu$ and the coproduct
$\Delta_0$. The above equation implies that
\begin{equation}\label{6.10}
\partial_\mu m_\phi = m_\phi \Delta_\phi \partial_\mu
\end{equation}
where $\Delta_\phi = \mathcal{F}^{-1}_\phi \Delta_0 \mathcal{F}_\phi$ and $\mathcal{F}_\phi$ is the twist
operator in $\phi$-realization. It follows from Eqs. \eqref{6.07} and \eqref{6.10} that
\begin{equation}\label{6.11}
m_\phi \Delta_\phi \partial_\mu \left(e^{ikx}\otimes e^{iqx}\right) = iD_\phi(k,q)_\mu\, e^{ikx}\star_\phi e^{iqx},
\end{equation}
where $D_\phi(k,q)_\mu$ denotes the $\mu$-component of $D_\phi(k,q)$. The coproduct $\Delta_\phi \partial_\mu$ has a generic form
\begin{equation}
\Delta_\phi \partial_\mu =\sum_{\a} A^{(1)}_{\a\mu}(\partial) \otimes A^{(2)}_{\a\mu}(\partial),
\end{equation}
thus using the above expression we find
\begin{equation}\label{6.12}
m_\phi \Delta_\phi \partial_\mu \left(e^{ikx}\otimes e^{iqx}\right) = \left(\sum_{\a} A^{(1)}_{\a\mu}(ik)\, A^{(2)}_{\a\mu}(iq)\right)
\, e^{ikx}\star_\phi e^{iqx}.
\end{equation}
Therefore, comparing Eqs. \eqref{6.11} and \eqref{6.12} we conclude
\begin{equation}\label{6.13}
D_\phi(k,q)_\mu = -i\widetilde \Delta_\phi (ik,iq)_\mu,
\end{equation}
where we have denoted
\begin{equation}
\widetilde \Delta_\phi (ik,iq)_\mu = \sum_\a A^{(1)}_{\a\mu}(ik)\, A^{(2)}_{\a\mu}(iq).
\end{equation}
Thus, Eq. \eqref{6.13} gives a correspondence between the coproduct $\Delta_\phi$ in
$\phi$-realization and the function $D_\phi$. Using this relation the star product in
$\phi$-realization may be written as
\begin{equation}\label{6.14}
(f\star_\phi g)(x) = m_0\Big\{e^{x(\Delta_\phi -\Delta_0)\partial^u}\, f(u)\otimes g(u)\Big\}_{u=x}.
\end{equation}
In the right, symmetric left-right and Weyl realizations Eq. \eqref{6.14} agrees with the
corresponding star-products in \cite{36}.

Let us find a relation between the star-products in different realizations. Let $T_S$ be the
similarity transformation which maps the $\phi_1$-realization to $\phi_2$-realization. Recall that
$S$ is explicitly given by $S=S_2 S_1^{-1}$ where $S_i$ is of the form \eqref{S}, and $T_{S_i}$
maps the right realization to $\phi_i$-realization. Fix $\hat f(X)\in U(\g)$, and let $X_\mu^{(i)}$
denote $X_\mu$ in $\phi_i$-realization. Then $\hat f(X^{(i)}) |0\rangle = f_i(x)$. Since $X_\mu^{(2)}=
SX_\mu^{(1)}S^{-1}$ we have
\begin{equation}
\hat f(X_\mu^{(2)}) = S \hat f(X^{(1)}) S^{-1} |0\rangle = S\hat f(X^{(1)}),
\end{equation}
where we have used $S^{-1}|0\rangle = 1$. Therefore, $f_2(x)=Sf_1(x)$ which implies that the
star-products in two realizations are related by
\begin{equation}
f\star_{\phi_2}g = S\left(S^{-1}f\star_{\phi_1}S^{-1}g\right).
\end{equation}

Using Eq. \eqref{6.13} one can deduce the function $D_\phi =(D_\phi^{(+)},D_\phi^{(-)},D_\phi^{(z)},D_\phi^{(\bar z)})$
for the NC space $NW_4$ from the coproduct $\Delta_\phi \partial_\mu$ found in Sec. 6:
\begin{align}
D_\phi^{(+)}(k,q) &= k_+ + q_+ \notag \\
&+\theta \frac{k_z k_{\bar z}}{\varphi(ak_-) \varphi(-ak_-)}\Big[F(ak_-)-F(ak_- + aq_-)\Big] \notag \\
&+\theta \frac{q_z q_{\bar z}}{\varphi(aq_-)\varphi(-aq_-)}\Big[F(aq_-)-F(ak_- + aq_-)\Big]  \notag  \\
&-\theta e^{-ak_-}\frac{k_z q_{\bar z}}{\varphi(ak_-)\varphi(-aq_-)}\Big[1+F(ak_- + aq_-)\Big]  \notag \\
&+\theta e^{ak_-}\frac{k_{\bar z} q_z}{\varphi(-ak_-)\varphi(aq_-)}\Big[1-F(ak_- + aq_-)\Big], \\
D_\phi^{(-)}(k,q) &= k_- + q_-, \\
D_\phi^{(z)}(k,q) &= \varphi(ak_- + aq_-)\left[\frac{k_z}{\varphi(ak_-)}+e^{ak_-}\frac{q_z}{\varphi(aq_-)}\right], \\
D_\phi^{(\bar z)}(k,q) &= \varphi(-ak_- -aq_-)\left[\frac{k_{\bar z}}{\varphi(-ak_-)}+e^{-ak_-}\frac{q_{\bar z}}{\varphi(-aq_-)}\right].
\end{align}

Deformed addition in $\phi$-ordering of the momenta $k$ and $q$ is defined by
\begin{equation}
k\oplus_{\phi} q = D_{\phi}(k,q).
\end{equation}
The binary operation $\oplus_{\phi}$ depends on the $\phi$-realization and represents a deformation of ordinary addition since
\begin{equation}\label{6.08}
k\oplus_{\phi}q = k+q+O(a,\theta).
\end{equation}
This nonabelian operation is associative, which follows from Eq. \eqref{6.07} and associativity of the star product.
The neutral element is $0\in \R^n$ since
\begin{equation}
k\oplus_{\phi}0 = D_\phi(k,0)=K^{-1}_\phi\left(D_s(K_\phi(k),0)\right) = K^{-1}_\phi (K_\phi(k))=k,
\end{equation}
and similarly $0\oplus_\phi k=k$. The inverse element, denoted $\underline{k}$, satisfies
\begin{equation}
k\oplus_{\phi}\underline{k} = \underline{k}\oplus_{\phi} k = 0.
\end{equation}
It follows from Eq. \eqref{6.08} that
\begin{equation}
\underline{k} = -k+O(a,\theta),
\end{equation}
hence $\underline{k}$ is a deformation of the ordinary opposite element $-k$, and it is the antipode of $k$.
The inverse element $\underline{k}$ can be found from the
condition $D_\phi(k,\underline{k})=0$:
\begin{equation}
\underline{k} = \left(-k_+,-k_-,-k_z\, e^{-ak_-}\frac{\varphi(-ak_-)}{\varphi(ak_-)},-k_{\bar z}\, e^{ak_-}\frac{\varphi(ak_-)}
{\varphi(-ak_-)}\right).
\end{equation}
In view of Eq. \eqref{6.07} we have
\begin{equation}
e^{ikx}\star_\phi e^{i\underline{k}x}=1.
\end{equation}
Note that in the Weyl realization when $\varphi$ is given by Eq. \eqref{W1} we have $\underline{k}=-k$. In fact,
the same is true for any function of the form $\varphi(k)=e^{-\frac{k}{2}} \widetilde \varphi(k)$ where $\widetilde \varphi$
is an even function.

We conclude this section by giving the explicit form of the twist operator $\mathcal{F}_\phi$ in
two special cases: (i) $\theta=0$, $a\neq 0$ and (ii) $a=0$, $\theta\neq 0$. When $\theta=0$ and
$a\neq 0$ the twist operator is given by
\begin{align}
\mathcal{F}_\phi &= \exp\Big[(\bar z\pzb \otimes 1)\ln\frac{\varphi(A_1+A_2)}{\varphi(A_1)}+
(z\pz \otimes 1)\ln\frac{\varphi(-A_1-A_2)}{\varphi(-A_1)} \\
&+ (1\otimes \bar z\pzb)\Big(A_1+\ln\frac{\varphi(A_1+A_2)}{\varphi(A_2)}\Big)
+ (1\otimes z\pz)\Big(-A_1+\ln\frac{\varphi(-A_1-A_2)}{\varphi(-A_2)}\Big)\Big], \notag
\end{align}
where we have denoted $A_1=A\otimes 1$ and $A_2=1\otimes A$, as before. Recall that $\varphi(A)=\exp((\gamma_0-1)A)$.
Hence, in the right realization $(\gamma_0=1)$ the operator $\mathcal{F}$ has a simple form
\begin{equation}
\mathcal{F}_R = \exp\big[A\otimes (\bar z\pzb -z\pz)\big].
\end{equation}
The twist operators for $\gamma_0=0$ and $\gamma_0=1$ were previously constructed in \cite{25}, and
are given by Eq. (60) and Eq. (61) respectively. For $\gamma_0=1/2$ the twist operator agrees with
the twist proposed by Bu et. al. in \cite{Bu}.

On the other hand, when $a=0$ and $\theta\neq 0$ we find
\begin{equation}
\mathcal{F} = \exp\left[\frac{\theta}{2}\big(x_+ \pz\otimes \pzb + \pz\otimes x_+ \pzb
-x_+ \pzb \otimes \pz - \pzb \otimes x_+ \pz\big)\right].
\end{equation}
In all cases the twist operator satisfes the cocycle condition \cite{49}, \cite{50}
\begin{equation}
(\mathcal{F}\otimes 1)(\Delta \otimes id)\mathcal{F} = (1\otimes \mathcal{F}) (id\otimes \Delta) \mathcal{F}.
\end{equation}
For kappa-deformed spaces in $n$ dimensions this relation was proved in \cite{GGHMM}. We remark
that the kappa-deformation of Poincare symmetries cannot be described by Drinfeld twist as an
element of the tensor product of two enveloping Poincare algebras. The twist operator considered
here is embedded in a larger algebra $U(igl(4))\otimes U(igl(4))$ where $igl(4)$ is the
inhomogeneous general linear algebra.

\section{Concluding remarks}

We have investigated a generalized kappa-deformed space of Nappi-Witten type which is a unification of kappa and
theta-deformed spaces with arbitrary deformation parameters $a$ and $\theta$. We have constructed an infinite
family of realizations of this space in terms of commutative coordinates $x_\mu$ in Euclidean space
and the corresponding derivatives $\partial_\mu$. All realizations are related by a group of similarity transformations
defined by the operator \eqref{S}. In particular, we have investigated a class of realizations \eqref{2.4}-\eqref{2.7}
parameterized by functions $\varphi$ and $F$. To each realization we have associated a corresponding
ordering prescription given by Eq. \eqref{K0}. For a special choice of $\varphi$ and $F$, and with $a=\theta$,
we reproduce the time ordering, symmetric time ordering and Weyl ordering constructed by Halliday and Szabo
\cite{36}. Unlike \cite{36}, in our approach the ordering prescriptions follow from a general procedure
and they are all related by similarity transformations.

Furthermore, we have extended the space of NC coordinates by introducing generalized derivatives.
We have shown that to each realization of the NC coordinates one can associate an extended phase
space which is a deformed Heisenberg algebra. The simplest extension is obtained in the right
realization when the extended algebra $\h$ is of Lie type, and it agrees with the deformed
Heisenberg algebra discussed in \cite{36}. The algebra $\h$ was further extended by introducing
rotation operators $M_{\mu\nu}$ which satisfy the ordinary $so(4)$ algebra. The coproduct $\Delta
M_{\mu\nu}$ is not closed in the tensor product of the enveloping algebras of $so(4)$. Hence, if
one wishes to have the coalgebra structure then it is natural to consider a larger algebra $igl(4)$
\cite{Bu}.

The Leibniz rule and coproduct was found for the extended algebra $\h$. Applying the similarity
transformations to this coproduct we derived the Leibniz rule and coproduct for all realizations
described by $\varphi$ and $F$. In the right realization, symmetric left-right and Weyl realization
this coproduct agrees with the coproducts considered in \cite{36} when $a=\theta$. We derived a
general formula for the star-product in terms of the coproduct, and we found an explicit expression
for the star-product in all realizations parameterized by $\varphi$ and $F$. In the above mentioned
special realizations these results reproduce the star-products found in \cite{36}. Also, we have
found a general form of the Drinfel'd twist operator for special values of the deformation
parameters: (i) $a\neq 0$, $\theta=0$, and (ii) $a=0$, $\theta\neq 0$. The twist operator is
embedded in the tensor product of two enveloping algebras of $igl(4)$.

The results in this paper are easily generalized to higher dimensionswhen the NC space under consideration
is generated by $X_+,X_-$ and $n$ copies of $Z$ and $\bar Z$ (see Eqs. \eqref{2.33}-\eqref{2.35}). In this
case the Weyl algebra $\mathcal{A}_4$ is replaced by $\mathcal{A}_{2n+2}$ and all expressions involving
realizations, ordering prescriptions, Leibniz rules, coproducts and twist operators should be modified
as follows. If $z,\bar z,\pz$ and $\pzb$ appear linearly, they are replaced by $z_\mu, \bar z_\mu,
\partial_{z_\mu}$ and $\partial_{\bar z_\mu}$. Any quadratic combinations involving $z\pz, \bar z\pzb$ and $\pz \pzb$ are replaced
by sums over $\mu$. Finally, we remark that all the results obtained here can be easily extended to
Mikonwski space.

Our general formalism can be further developed in order to construct and analyize QFT on NC spaces
generalizing the results obtained by Halliday and Szabo \cite{36} by applying twist operators in a systematic
way. These problems will be addresed in future work.

\textit{Aknowledgements.} The authors wish to thank Z. \v{S}koda and M. Dimitrijevi\'{c} for useful
discussions and comments. This work is supported by the Croatian Ministry of Science, Education and
Sports grants no. 098-0000000-2865 and 177-0372794-2816.

%%%%%%%%%%%%%%%%%%%%%%%%%%%%%%%%%%%%%%%%%%%%%%%%%%%%%%%%%%%%%%%%%%%%%%%%%%%%%%%%%%%%%%%%%%%%%%%%%%%%%%%%%%%%%%

%\section*{References}

\end{document}